\documentclass[useAMS,usenatbib]{mn2e}
\usepackage{graphicx}    % needed for including graphics e.g. EPS, PS
\usepackage{subfigure}
\usepackage[latin1]{inputenc}
\usepackage{amsmath}
\usepackage{amsfonts}
\usepackage{amssymb}

\def \fps@figure{htbp}
\makeatother

\DeclareGraphicsExtensions{.ps,.pdf,.png}
\makeatletter

\def \sersic  {S\'{e}rsic}

\def \re      {$R_{\rm e}$}

\def \vdvir   {$\sigma_{200}$}
\def \Mvir    {$M_{200}$}

\begin{document}

\title[Evolution of the Brightest Cluster Galaxies]{Evolution of the Brightest Cluster Galaxies: the influence of morphology, stellar mass and environment}

\author[Zhao et al.]{Dongyao Zhao$^{1}$\thanks{E-mail: : \texttt{ppxdz1@nottingham.ac.uk}}, 
Alfonso Arag\'{o}n-Salamanca$^{1}$\thanks{E-mail: : \texttt{alfonso.aragon@nottingham.ac.uk}}, 
Christopher~J.~Conselice$^{1}$\thanks{E-mail: : \texttt{christopher.conselice@nottingham.ac.uk}}\\
\footnotemark[0]\\$^{1}$School of Physics and Astronomy, The University of Nottingham, University Park, Nottingham, NG7 2RD, UK}

\date{Accepted ??. Received ??; in original form ??}
\pagerange{\pageref{firstpage}--\pageref{lastpage}} \pubyear{2014}
\maketitle

\label{firstpage}

\begin{abstract}
Using a sample of $425$ nearby Brightest Cluster Galaxies (BCGs) from \cite{Linden07}, we study the relationship between their internal properties (stellar masses, structural parameters and morphologies) and their environment. More massive BCGs tend to inhabit denser regions and more massive clusters than lower mass BCGs. Furthermore, cDs, which are BCGs with particularly extended envelopes, seem to prefer marginally denser regions and tend to be hosted by more massive halos than elliptical BCGs. cD and elliptical BCGs show parallel positive correlations between their stellar masses and environmental densities. However, at a fixed environmental density, cDs are, on average, $\sim40$\% more massive. Our results, together with the findings of previous studies, suggest an evolutionary link between elliptical and cD BCGs. We suggest that most present-day cDs started their life as ellipticals, which subsequently grew in stellar mass and size due to mergers. In this process, the cD envelope developed. The large scatter in the stellar masses and sizes of the cDs reflects their different merger histories. The growth of the BCGs in mass and size seems to be linked to the hierarchical growth of the structures they inhabit: as the groups and clusters became denser and more massive, the BCGs at their centres also grew. This process is nearing completion since the majority ($\sim 60$\%) of the BCGs in the local Universe have cD morphology. However, the presence of galaxies with intermediate morphological classes (between ellipticals and cDs) suggests that the growth and morphological transformation of some BCGs is still ongoing. 
\end{abstract}


\section{Introduction} 
The brightest cluster galaxies (BCGs) are the most luminous and massive galaxies in the universe. They are found at the centres of galaxy clusters and groups, and exhibit many unique properties  \citep[see, e.g.,][]{Tonry87,KD89,Jordan04,Linden07}. Their origin and evolution is intimately linked with the evolution of their host clusters, and therefore can provide direct information on the formation and history of large-scale structures in Universe (\citealt{Conroy07}).

Many scenarios have been proposed to explain the formation and evolution of BCGs. One of them is galactic cannibalism \citep{White76,OH77,Garijo97}, where BCGs were formed as a result of hierarchical mergers of smaller galaxies. Other hypotheses include tidal stripping from cluster galaxies \citep{Richstone76,Merritt85}, and star formation in the cluster core, where BCGs are formed through cooling flows \citep{Fabian94}. Recently, numerical simulations and semi-analytic models suggest a two-phase process for BCGs formation. In these models, the stellar component of BCGs was initially formed through the collapse of cooling gas or gas-rich mergers at high redshifts; subsequently, BCGs continued to grow substantially by dissipationless processes such as dry mergers \citep{DeLB07,Naab09,Laporte12}. This inside-out formation scenario is broadly consistent with observations, avoiding the need for cooling flows to provide the cold gas that would be necessary if BCGs had formed at later times. It also overcomes the problem caused by the merger rate in clusters being too low due to the high velocity dispersion in dynamically relaxed clusters. However, some studies such as \citet{Ascaso11} claimed that feedback rather than merging processes are the main mechanism affecting the evolution of the BCGs to the present epoch, ending the star formation within these systems. Therefore, many important details in the processes governing BCG formation and evolution are still unclear and deserve further investigation.

Since BCGs posses singular properties (e.g., distinct structures and morphologies, and very high stellar masses) and reside in special environments (the core of groups and clusters), studying the relationship between their properties and their environments will help to constrain the theories of BCG formation and evolution and tell us whether the intrinsic properties of BCGs or the environment play a dominant role in their history. In this context, it is important to bear in mind that, while both the location of BCGs at the bottom of the potential wells of clusters and their dominance at the massive end of the galaxy luminosity function may influence their properties, it is nonetheless very difficult to disentangle these two influences since it is hard to find equally massive non-BCGs for comparison. Therefore, when comparing BCGs and non-BCGs, differences in the mass range spanned by the samples may bias the results.

One key observational property of BCGs is that many of them show unique morphologies. The vast majority (but not all, see \citealt{Zhao15}) BCGs are early-type galaxies. Most BCGs are classified as either elliptical or cD galaxies \citep{LP92,Fasano10,Zhao15}. 
The defining characteristic separating these two morphological types is the presence of an extended, low-surface-brightness stellar envelope in cDs that is absent in ellipticals (e.g. \citealt{Dressler84}; \citealt{OH01}). Since cDs are not found outside the BCG galaxy population, it is very important to consider this unique galaxy class when studying BCGs. We will therefore use morphology as one of the main observables in this paper, focusing on the different properties of elliptical and cD BCGs. 

Many previous observational works usually study the BCG population as a whole, and compare it with the population of elliptical galaxies that are not BCGs \citep{Bernardi07,Lauer07,Linden07,Liu08}. However, there has been some recent work exploring the structural differences between cluster ellipticals and BCGs with different morphologies. \cite{Fasano10} found that, while non-BCG cluster ellipticals generally have triaxial shape with a weak preference for prolateness, BCGs are also triaxial but with a much higher tendency towards prolateness. Such a strong prolateness appears entirely due to the fact that cDs dominate the BCG population. In fact, while the shape of elliptical BCGs does not differ from other cluster ellipticals, cDs tend to have prolate shapes. Furthermore, they suggest that the prolateness of the cDs could reflect the shape of the associated dark matter haloes. More recently, \cite{Zhao15} have studied in detail the morphology and structure of BCGs, demonstrating that the morphological distinction between ellipticals and cDs is accompanied by quantitative structural differences. cD BCGs generally have much larger sizes and their light profiles cannot be modelled accurately using single \sersic\ functions. Conversely, elliptical BCGs are smaller and single \sersic\ profiles provide better fits to their surface brightness distributions. These differences in morphology and structure suggest that cD and elliptical BCGs have followed different evolutionary paths. We investigate these possible scenarios in this paper.

There has been a significant amount of work addressing the formation and evolution of BCGs.  For example, \citet{Guo09} studied how the structural parameters of central cluster galaxies correlate with their stellar masses and their host dark matter (DM) halo mass. They found that stellar mass is the dominant property dictating the shape and size of these galaxies, and suggest that the DM halo mass does not play a very significant role. \citet{Hogg04}, \citet{Kauffmann04} and \citet{vanderWel08} also reached similar conclusions. In contrast, other studies (e.g., \citealt{Ascaso11}) claimed that there is a significant correlation between the cluster mass and the properties of BCGs. Furthermore, \citet{Tovmassian12} added the cluster richness to the halo/cluster mass as another environmental indicator. They found that the absolute $K$-band luminosity of cD galaxies (a good proxy for stellar mass) strongly depends on the cluster richness, but less strongly on the cluster velocity dispersion (a proxy for DM halo mass). Therefore, since the effects of the halo mass and the cluster richness could be different, it is necessary to take them into account as separate environmental parameters when studying BCG evolution. 

Many other recent papers have studied the properties of BCGs in relation to other early-type galaxies, providing important clues to how they form and evolve. Some examples  include \cite{Shankar2013,Shankar2014a,Shankar2014b,Shankar2015,HC2013a,HC2013b,Bernardi2009}. For the sake of brevity, we will not describe their findings here but we will mention them in the following discussion when relevant.  

In this paper we use a well-defined local sample of $625$ BCGs from \citet[][hereafter L07]{Linden07} and carry out a comprehensive and systematic statistical study on the correlation between BCGs intrinsic properties (structure, morphology and stellar mass) and their environment. We consider two environmental measures, a global one (the DM cluster halo mass, characterised by its velocity dispersion) and a local one (the galaxy density). In doing so we will obtain very valuable additional information on how BCGs form and evolve.  

The galaxy groups and clusters these BCGs inhabit span a very broad range of total masses, from $\sim10^{13}\,M_\odot$ to
$\sim10^{15}\,M_\odot$. Since there is no clear boundary separating ``clusters'' from ``groups'' (although $10^{14}\,M_\odot$ could be taken as the transition mass), we will study group and cluster BCGs together. We will explore how the masses of the parent groups/clusters affect the properties and evolution of the BCGs.  

The paper is organized as follows. In \S\ref{sec:data} we introduce the BCG sample, and describe the observables we will use (morphologies, structural parameters, stellar masses, environmental densities, and DM halo virial masses). In \S\ref{sec:earlytype} we show how the structural parameters of the BCGs relate to their stellar masses, and their global and local environment, and discuss the implications of the correlations we find on the formation of the BCG population. In \S\ref{sec:cDBCGs} we go one step further and bring the galaxy morphologies into the general picture to learn about the distinct evolutionary history of cD and elliptical BCGs. We summarise our main conclusions in \S\ref{sec:conclude}. Throughout this paper we have adopted the $\Lambda$CDM cosmology with $\Omega_{\rm m}=0.3$, $\Omega_{\Lambda}=0.7$, and $H_0=70$ km s$^{-1}$ Mpc$^{-1}$.

\section{BCG sample and properties}
\label{sec:data}

The parent BCG sample we use in this paper comes from the catalogue published by L07. The groups and clusters that host these BCGs are contained in the SDSS-based C4 cluster catalogue \citep{Miller05}, a widely-used and well-defined sample whose reliability has been thoroughly tested by simulations. Based on the C4 sample, L07 developed an improved algorithm to identify the BCG in each cluster and published a catalogue containing $625$ BCGs residing in galaxy groups and clusters at $0.02\leqslant z \leqslant 0.10$. See \cite{Linden07} for a detailed discussion on the BCG identification method.

In our previous paper \citep{Zhao15} we published visual morphologies for these $625$ BCGs. The BCGs were classified into three main types: 414 cDs, including pure cD (356), cD/E (53) and cD/S0 (5); 155 ellipticals, including pure E (80), E/cD (72), and E/S0 (3); 46 disk galaxies, containing spirals (24) and S0s (22). There are also 10 BCGs undergoing major mergers. We used intermediate classes such as cD/E (probably a cD, but could be E) and E/cD (probably E, but could be cD) to account for the uncertainty inherent in the visual classification. Separating cD BCGs and non-cD elliptical BCGs is a very hard problem since there is no sharp morphological distinction between these two classes \citep[e.g., ][]{patel06,Liu08}. Detecting the extended stellar envelope that characterises cD galaxies depends not only on its dominance, but also on the quality and depth of the images, and on the details of the classification method(s) employed. Since the SDSS imaging data that we use is of uniform quality, the classification is internally consistent. Moreover, \cite{Zhao15} demonstrated that the morphological type of a galaxy is very tightly related to its structural parameters, and devised a quantitative method to separate cDs from elliptical BCGs in a robust manner that agrees very well with the visual classification. We are therefore confident that the morphological information that we use here is reliable and self-consistent within our sample. Obviously, a degree of caution would be necessary when comparing our morphologies with those of galaxies from other samples since the quality of the images and the classification criteria may be different.   However, in \cite{Zhao15} we carried out imaging simulations to show that the outer envelopes of cD galaxies would have been detected if present, and thus that there is no bias in our morphological distinction between E and cD.

In this paper we will call ``cD BCGs'' the 414 galaxies classified by \cite{Zhao15} as cD, cD/E and cD/S0, and ``elliptical BCGs'' the 155 galaxies  classified as E, E/cD and E/S0. We will also include in our study the 46 disk BCGs (spirals and S0s), but not the 10 major mergers.  This sample therefore contains 615 BCGs.

The structural properties (\sersic\ index $n$ and effective radius \re)\footnote{Strictly speaking, \re\ is the effective semi-major axis of the single \sersic\ model fit} that we use in this paper were also published by \citet{Zhao15}. These were derived from SDSS DR7 $r$-band images using two-dimensional single \cite{Sersic63} model fits to the galaxies' light profiles. 
The fits were carried out with GALFIT \citep{Peng02} using the GALAPAGOS \citep{Barden12} pipeline. The method simultaneously fits the target galaxy and its near neighbours, yielding more accurate fits and improved sky subtraction. The imaging data reaches a surface brightness limit of $\sim27\,$ mag/arcsec$^{2}$, and are therefore deep enough to study the faint extended envelopes present in cD BCGs. The detailed description of the fitting procedure and structural parameter estimation can be found in \citet{Zhao15}. The values of \re\ and $n$ that we obtained are broadly compatible with the ones published by \cite{Guo09}. However, there are some relatively minor systematic differences due to the improvements in the sky subtraction procedure implemented by \citet{Zhao15}. A direct comparison is presented in Appendix~\ref{sec:guo1}.  

The stellar masses we use come from ``The MPA--JHU DR7 release of spectrum measurements'' (see \texttt{www.mpa-garching.mpg.de/SDSS/DR7/})\footnote{In this paper we use their updated stellar masses from \texttt{http://home.strw.leidenuniv.nl/$\sim$jarle/SDSS/}}. Hereafter we call these ``MPA--JHU masses''. These stellar masses are obtained via spectral energy distribution (SED) fits to the DR7 photometric data using a \cite{Kroupa01} Initial Mass Function. Although the method is not identical to that of \citet{Kauffmann03} or \citet{Gallazzi05}, who use spectroscopic information, the resulting masses agree very well with only a few minor offsets. A detailed discussion and comparison of the methods can be found in \texttt{www.mpa-garching.mpg.de/SDSS/DR7/mass\_comp}. The number of BCGs in our sample which have MPA--JHU stellar mass information is $591$, i.e., $96$\%. The very small minority of galaxies without stellar masses include $20$ galaxies for which no spectroscopic redshift is available (essential to determine accurate distances) and $4$ for which the MPA--JHU catalogue fails to provide a value for the mass, presumable because the SED fitting method does not yield a reliable solution. Since only $4$\% of the galaxies in the parent sample do not have stellar masses, we do not expect them to have any significant influence in our results. At this stage, and in order to ensure we have a stellar-mass-selected sample, we impose a minimum mass of $3\times10^{10}\,$M$_{\odot}$, which reduces the sample to $535$ BCGs. This limit also eliminates a few galaxies whose stellar masses, structural parameters and morphologies have larger uncertainties due to their faint magnitudes. 

These MPA--JHU stellar masses are derived from Petrosian magnitudes and are therefore not dependent on the fitting parameters that we obtain. This is important since it allows us to look for independent correlations between stellar mass and the fit parameters. Alternatively, \cite{Guo09} estimated stellar masses using photometric fluxes derived from their light profile model fits. Such a method results in model-dependent stellar masses, which may produce spurious correlations between the masses and the model parameters. We will discuss this in more detail in section~\ref{sec:earlytype}, and we will argue that for our study the MPA--JHU Petrosian-based stellar masses should be preferred. 

The final key ingredients in our study are quantitative measurements of the environments where the BCGs reside. We will use two distinct descriptions of the environment, global  and local. The ``global environment'' is governed by the properties of the cluster/group that contains the BCG, and in particular its total mass (including the dark-matter halo). We use the velocity dispersion of the cluster (\vdvir) published by L07 to estimate the halo virial mass \Mvir\ using the Equation~10 of \cite{finn05}, which is
\begin{equation}
\begin{split}
M_{200} =   1.2\times10^{15}&\left(\frac{\sigma_{200}}{1000 \rm\,km\,s^{-1}}\right)^3 \hfill\  \\
 & \quad \times\frac{1}{\sqrt{\Omega_\Lambda+\Omega_0(1+z)^3}}\,h^{-1}_{100}\,M_\odot.
\end{split}
\end{equation}
The group and cluster sample studied here covers a broad range of masses, from $M_{200}\sim10^{13}\,M_\odot$ to 
$M_{200}\sim10^{15}\,M_\odot$, peaking at $M_{200}\sim10^{14}\,M_\odot$ (see Fig.~\ref{fig:cDEparm}).

To characterise the ``local environment'' we use the environmental luminosity density  introduced by \cite{Tempel12}. This is a good proxy for the environmental stellar mass density, which, as argued by \cite{wolf09}, is a better and more robust measurement of the environment than galaxy number density. The main advantages of using stellar mass (or luminosity) density over galaxy number density are twofold. First, the environmental luminosity/mass density does not depend strongly on the exact details of the galaxy sample used to define it, such as the magnitude limit, provided that it reaches significantly fainter than the ``knee'' of the luminosity function. And second, it represents better the strength of the interactions that a galaxy may experience from its neighbours: it is not the same to be surrounded by $N$ faint low-mass galaxies than by $N$ bright high-mass ones. \cite{Tempel12} determined these environmental densities using SDSS $r$-band luminosities with a smoothing scale of $1\,$h$^{-1}$Mpc. The total number of BCGs in our mass-limited sample for which we have both stellar masses and environmental densities is  $425$. The galaxies for which environmental densities are not available are outside the footprint of the contiguous sky region covered by the work of  \cite{Tempel12}, and therefore there is no reason to believe that their exclusion from our analysis will bias our conclusions. The BCG sample covers one order of magnitude in environmental density (see Fig.~\ref{fig:cDEparm}).

In what follows, we will consider the sample comprising the $425$ $M_*>3\times10^{10}\,$M$_\odot$ BCGs with cD (275), elliptical (116), S0 (15) and spiral (19) morphologies for which we have obtained stellar masses, cluster masses and environmental densities.

\section{Correlations between BCG properties}
\label{sec:earlytype}

In this section we analyse the correlations (or lack thereof) between the structural parameters, masses and environments (global and local) of the BCG population as a whole and discuss their implications. In section~\ref{sec:cDBCGs} we will include morphology as an additional key property.

\subsection{Stellar Masses and Structural Parameters}
\label{sec:stellarmass}

\begin{figure}
\centering{ %\centering
\subfigure{
\centering{\includegraphics[scale=0.66]{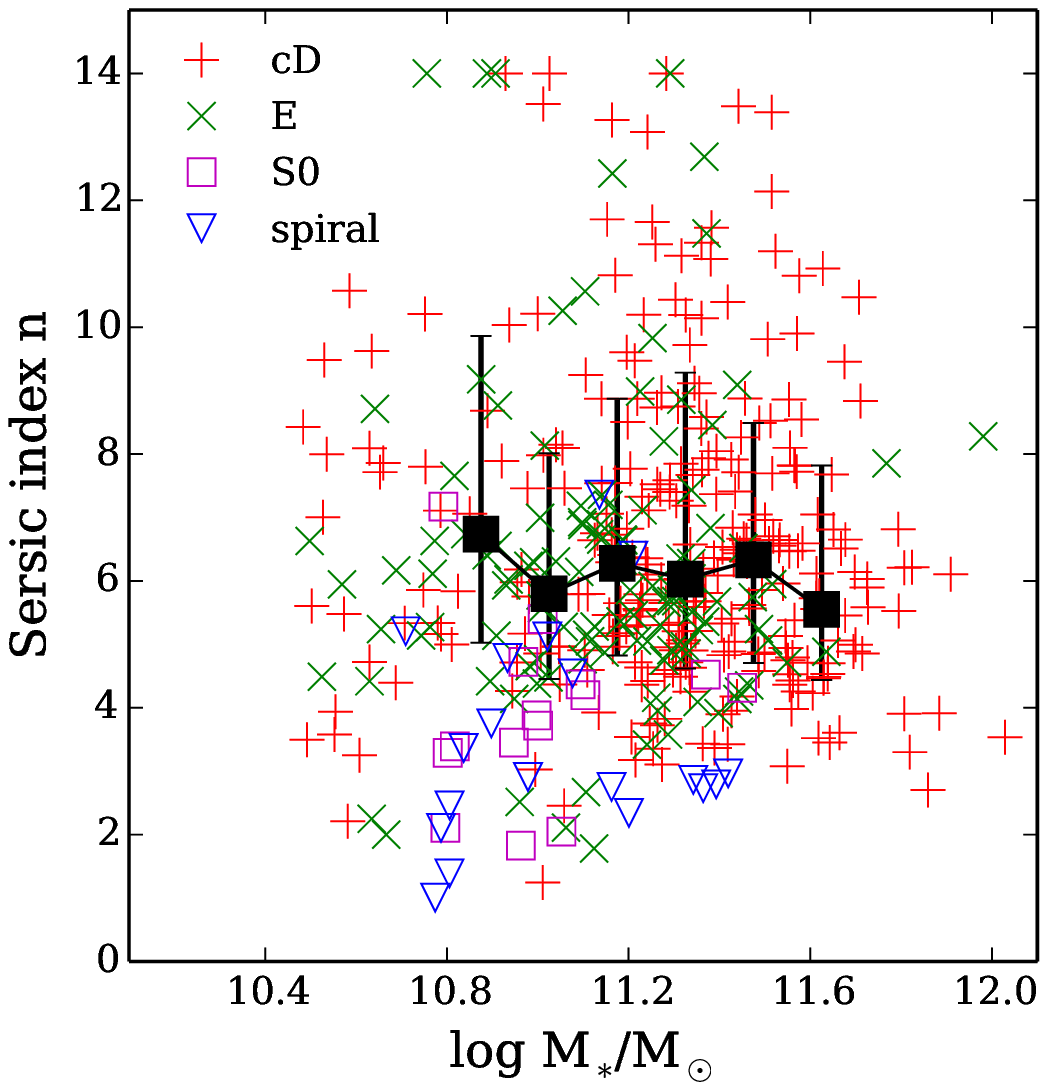}}}  
\subfigure{
\centering{\includegraphics[scale=0.64]{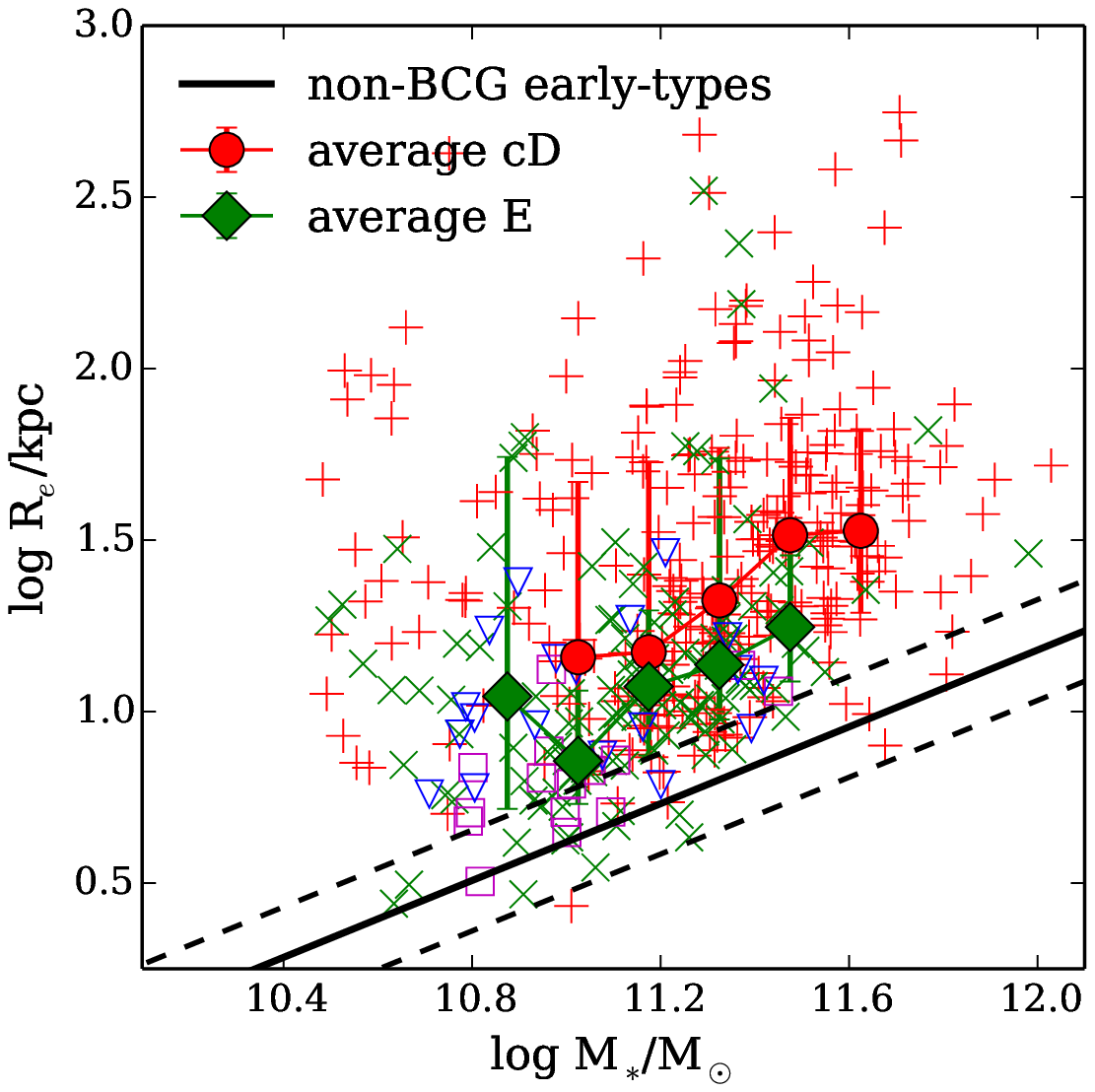}}}
\caption{Comparison between the stellar masses and the structural parameters of the BCGs in our sample. Upper panel: \sersic-index $n$ vs.\ MPA--JHU stellar mass $M_*$.  Lower panel: effective radius \re\ vs. $M_*$. Red plus signs, green crosses, magenta open squares and blue open triangles correspond to cD, elliptical, S0 and spiral BCGs, respectively. Black solid squares with error bars in upper panel show the median and the 84 and 16 percentiles ($\sim1\sigma$) of each parameter in $0.15\,$dex $\log M_{*}$ bins for the combined cD and elliptical BCGs. Red dots and green diamonds with error bars in lower panel are for cD and elliptical BCGs, respectively. Bins with fewer than $20$ galaxies are excluded due to their large statistical uncertainties. The black solid line in the lower panel corresponds to the best-fit relation for the normal (non-BCG) early-type galaxy population, defined to have $n>2.5$, from \citet{Shen03}. The dashed lines correspond to the $1\sigma$ scatter in this relation. 
\label{fig:nrem}}}
%\vspace{-0.2cm}
\end{figure}

First we explore the relation between the BCGs structural parameters (\sersic\ index $n$ and effective radius \re) and their stellar mass $M_*$. In the top panel of Fig.~\ref{fig:nrem} we investigate whether there is a statistical correlation between the galaxies' profile shape, characterised by $n$, and their stellar mass. To guide the eye, we have binned the data in stellar mass bins $0.15\,$dex wide. The black squares with error bars show the median and the $84$ and $16$ percentiles ($\sim 1\sigma$) of the $n$ distributions for each mass bin, considering only the BCGs with cD and elliptical morphologies. In order to avoid large statistical uncertainties, we exclude bins with fewer than $20$ galaxies. 

We find no correlation between $n$ and $M_{*}$ for these galaxies. The median $n$ for the elliptical and cD BCGs is $6.02$, which indicates that, on average, these galaxies have both centrally-concentrated light profiles and extended envelopes, as expected for a population dominated by cDs \citep[see][and references therein]{Zhao15}. Interestingly, as \cite{Zhao15} pointed out, there is little separation between the $n$ distributions of cD and elliptical galaxies. A Kolmogorov-Smirnov test indicates that the difference is only significant at the $2\sigma$ level. The median \sersic\ index $n$ is $6.12^{+2.76}_{-1.63}$ for cDs and $5.86^{+2.31}_{-1.42}$ for ellipticals \citep{Zhao15}\footnote{The errors quoted for median values correspond to the 84 and 16 percentiles of the distributions ($\sim1\sigma$ scatter).}. The slightly larger median $n$ value of the cD galaxies is driven by their extended envelope. As expected, disk BCGs (spirals and S0s) have significantly lower $n$ values ($2.91$ and $3.88$ respectively).

The lack of correlation between $n$ and $M_*$ for the BCGs in our sample contrasts with the findings of \citet{Guo09}, who claimed a clear positive correlation in the sense that more massive BCGs seem to have higher values of $n$. As we show in Appendix~\ref{sec:guo2}, we believe this may be due to the fact that \citet{Guo09} estimated stellar masses from total luminosities derived from single \sersic\ model fits. These luminosities (and the derived stellar masses) depend on the value of $n$, and this dependency could drive an artificial correlation.

As an aside, we note that in the upper panel of Fig.~\ref{fig:nrem} there is a small number of cD and elliptical BCGs whose $n$ is quite large ($n>12$). It is important to realise that for large $n$ ($n>6$ or so) very small changes in the light profile result in large changes in $n$, and thus all values of $n$ above $\sim 6$ correspond essentially to the same profile. Furthermore, a visual inspection of the fits and the residuals indicate that these large $n$ objects are usually surrounded by multiple close bright companions (or, in a few cases, a bright nearby star). This makes the fits less reliable. Furthermore, some of these objects have double cores, and therefore a single \sersic\ profile is not a good model of their surface brightness distribution. In these cases, the derived model parameters should be taken with caution. Since the fraction of affected objects is quite small, they do not affect the statistical conclusions of this study. Removing them would have no significant statistical effect, and they are therefore kept in our analysis for completeness.   Another reason for this that the high $n$ systems are distributed over all stellar masses, and not just found within the high or low stellar mass systems.

We examine now the relationship between the effective radius \re\ and the stellar mass of the BCGs shown in the lower panel of Fig.~\ref{fig:nrem}. For comparison, we show the relation found for normal non-BCG early-type galaxies by \citet{Shen03} selected from the SDSS survey as system with $n > 2.5$. The sizes and stellar masses published by \citet{Shen03} are directly comparable to the ones we use. Their effective radii are computed from single \sersic\ fits to SDSS images, like ours, and their stellar masses are also derived using the method of \citet{Kauffmann03}. Note that the \citet{Shen03} sample is dominated by field galaxies, although we will see below that similar conclusions are obtained for cluster early-types. 

The effective radii of early-type BCGs is strongly correlated with their stellar masses: on average, \re\ increases when $M_*$ increases, but the scatter is large (about $\sim0.3\,$dex, or a factor of $\sim2$ in \re\ at a given mass). 
In agreement with \cite{Bernardi2009}, we find that almost all the BCGs are above the average relation for non-BCG early types, and the slope is similar (within a large uncertainty). The scatter is also larger for the BCGs than for the other early-type galaxies. Notwithstanding this large scatter, the median radius of BCGs is about twice as large as that of non-BCG early types of similar masses. This difference is largely due to the cD galaxies, which dominate the sample. As shown in the lower panel of Fig.~\ref{fig:nrem}, when we analyse the properties of BCGs separated by morphology, elliptical BCGs are, on average, significantly smaller than cDs. The minority of BCGs that have disk (spiral and S0) morphologies tend to populate the low end of the size distribution.

Fig.~\ref{fig:nrem} also shows that the BCGs in our sample span a very broad range of stellar masses ($10^{10.5}$--$10^{12}\,M_\odot$). This is mainly due to the fact that these BCGs are hosted by galaxy groups and clusters with very different masses (Fig.~\ref{fig:cDEparm}), combined with the weak correlation between the galaxies' stellar masse and \Mvir\ (Fig.~\ref{fig:haloparamt}). Nevertheless, it is clear that at all stellar masses BCGs have larger radii than non-BCG early-type galaxies.  This agrees with the findings of \cite{Valentinuzzi10} and \cite{Vulcani14} for low-redshift BCG and non-BCG galaxies in the WINGS clusters (see their Fig.~11). 
Although a detailed quantitative comparison is very difficult given the differences in methodology combined with the fact the the WINGS sample does not include groups, it is reassuring to see that compatible results are obtained independently. Note also that the stellar masses of the WINGS BCGs are all in the range $10^{11}$--$10^{12}\,M_\odot$, where most of our BCGs lie, but we also have BCGs with lower stellar masses since our sample includes both clusters and groups.

\subsection{Local environment: the effect of galaxy density}
\label{sec:edensity}

\begin{figure*}
\centering{\includegraphics[scale=0.72]{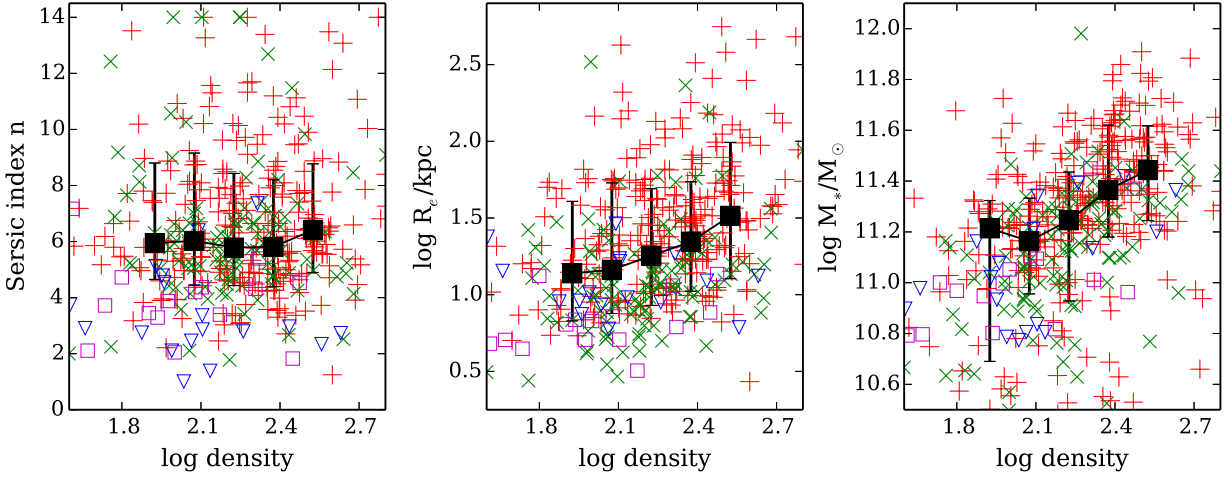}}
\caption{Relationship between environmental density and BCG properties. From left to right, these properties are the \sersic\ index $n$, the effective radius \re, and the stellar mass $M_*$. Symbols as in Fig.~\ref{fig:nrem}. 
\label{fig:envparamt}}
\end{figure*}

\begin{figure*}
\centering{\includegraphics[scale=0.72]{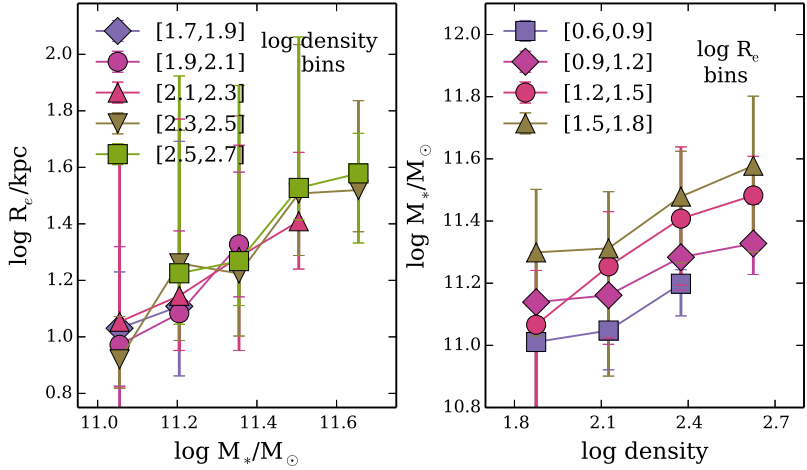}}
\caption{Left panel: \re\ vs.\ $M_{*}$ for cD and elliptical BCGs binned by environmental density. Right panel: $M_{*}$ vs.\ environmental density binned by \re. The points correspond to the median for each bin and error bars indicate the 84 and 16 percentiles ($\sim1\sigma$). Bins containing fewer than 5 galaxies have been excluded due to their large statistical uncertainties. The legend shows the different symbols corresponding to each bin. 
\label{fig:envparamt_bin}}
%\vspace{-0.2cm}
\end{figure*}

We explore now the relationship between the local environment that BCGs inhabit and their intrinsic properties (structural parameters and stellar masses). As discussed in Section~\ref{sec:data}, we use the environmental luminosity density of \cite{Tempel12} to characterise the local environment. In the three panels of Fig.~\ref{fig:envparamt} we plot the \sersic\ index $n$, the effective radius \re, and the MPA--JHU stellar mass $M_*$ vs.\ this density. The left panel shows that there is no correlation between $n$ and density (Pearson correlation coefficient $0.03$). However, both \re\ and $M_*$ clearly correlate, on average, with density (correlation coefficients $0.32$ and $0.49$ respectively). Although there is significant scatter, larger and more massive BCGs tend to inhabit in denser environments.

It appears that local density correlates with both the size and the stellar mass of the early-type BCGs. However, Fig.~\ref{fig:nrem} shows that \re\ correlates with $M_*$. It is therefore important to ascertain which of these two parameters is the intrinsic driver of the correlations with density. To do this, in the left panel of Fig.~\ref{fig:envparamt_bin} we plot \re\ vs.\ $M_{*}$ binning the galaxies by density. We only include cD and elliptical BCGs. For a given stellar mass, the median \re\ is the same for all densities. This  suggests that density does not affect BCG size directly, but only through its dependence with stellar mass. In the right panel of this figure we show the $M_*$--density relation again, but now binning the galaxies by radius. For galaxies of all sizes, there is a clear correlation between stellar mass and environment: more massive BCGs tend to inhabit denser regions, regardless of their radius. This implies that the stellar mass--density correlation is the more fundamental one, and that the environment affects the BCG stellar mass more directly than their sizes.

The fact that the mass-size relation for the general galaxy population does not depend significantly on environment (at least at low redshift) has been found in several recent studies \citep[e.g.,][]{Shen03,Maltby10,rettura10,HC2013a,HC2013b,poggianti13}. Our results reveal that this is also true for BCGs.

\subsection{Global environment: the effect of the cluster mass}
\label{sec:halomass}

\begin{figure*}
\raggedright{\includegraphics[scale=0.6]{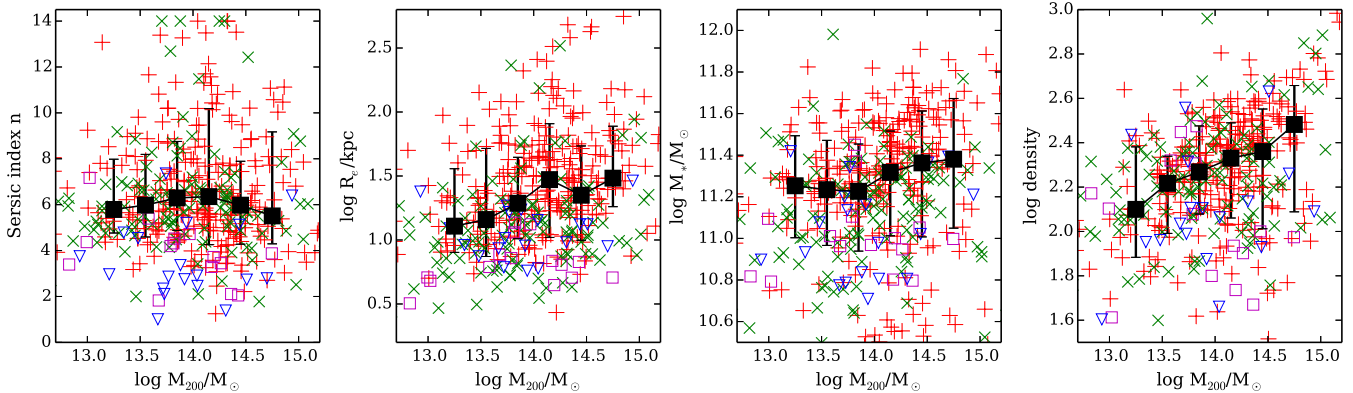}}
\caption{Relationship between \Mvir\ and other BCG properties. From left to right, these properties are the \sersic\ index $n$, the effective radius \re, the stellar mass $M_*$ and the environmental density. Symbols as in Fig.~\ref{fig:nrem}
\label{fig:haloparamt}}
%\vspace{-0.2cm}
\end{figure*}

\begin{figure*}
\raggedright{\includegraphics[scale=0.59]{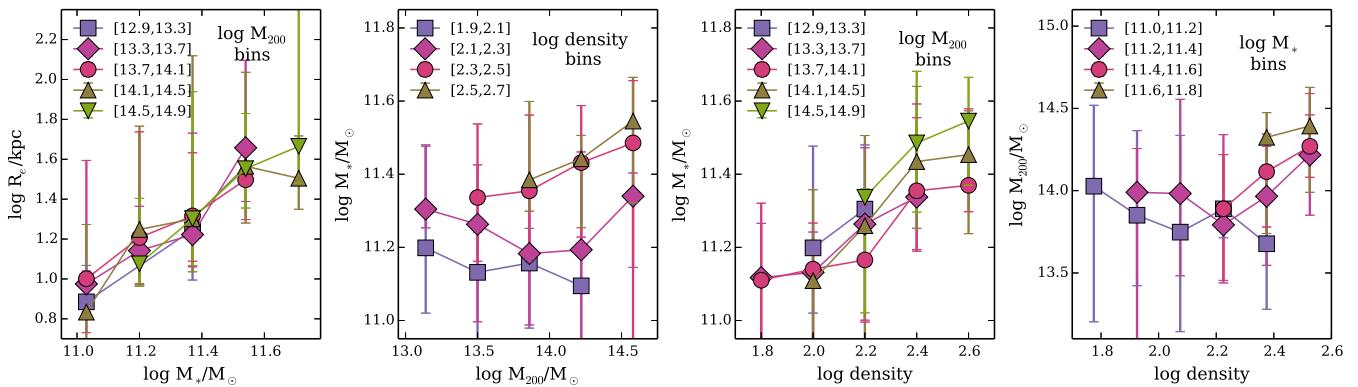}}
\caption{From left to right, the first panel shows \re\ vs.\ $M_{*}$ in \Mvir\ bins; the second panel $M_*$ vs.\ \Mvir\ in density bins; the third panel $M_*$ vs.\ density in \Mvir\ bins; and the fourth panel \Mvir\ vs. density in $M_*$ bins. The points correspond to the median for each bin and error bars indicate the 84 and 16 percentiles ($\sim1\sigma$). Bins containing fewer than 5 galaxies have been excluded due to their large statistical uncertainties. The legend shows the different symbols corresponding to each bin. Only cD and elliptical BCGs have been included.  
\label{fig:haloparamt_bin}}
%\vspace{-0.2cm}
\end{figure*}

We now consider the effect of the global environment (characterised by the total mass of the host cluster \Mvir; see Section~\ref{sec:data}) on the properties of the BCGs.  Fig.~\ref{fig:haloparamt} shows the relation of \Mvir\ with the \sersic\ index $n$, effective radius \re, stellar mass $M_*$ and environmental density (from left to right).   

The \sersic\ index does not show any dependence on the halo virial mass (Pearson correlation coefficient $-0.04$). Both effective radius and stellar mass show a small degree of correlation with \Mvir, albeit with large scatter (correlation coefficients $0.26$ and $0.17$ respectively)\footnote{Note that \re\ and $M_*$ correlate more weakly with \Mvir\ than with the environmental density (compare Figs.~\ref{fig:envparamt} and~\ref{fig:haloparamt}). 
%\cite{Shankar2014b} also found a correlation between $M_*$ and the parent halo mass, albeit with smaller scatter. 
}. As before, we need to explore which of these two parameters is the driver of the observed correlations. The first panel of Fig.~\ref{fig:haloparamt_bin} shows that the stellar mass--size relation does not depend on the \Mvir\ (global environment), in agreement with the findings of \cite{Shankar2014b}. Since we also found in Section~\ref{sec:edensity} that the size of BCGs is not directly affected by the local environment (or galaxy density) we conclude that any apparent environmental effect on \re\ is driven by the stellar mass--size relation combined with the environmental dependence (or dependencies) of stellar mass.

We now consider the effect of environment on the BCGs' stellar masses. Previous studies have found that the stellar masses of the BCGs correlate with the total mass (or velocity dispersion) of the host cluster \citep[e.g.,][]{Whiley08,Ascaso11}. One complication that plagues all environmental studies is the fact that the two characterisations of the environment that we use (local and global) are, not surprisingly, correlated (see rightmost panel of Fig.~\ref{fig:haloparamt}), although not very tightly (correlation coefficient $0.33$). However, these two measures of environment are clearly not representing the same physical scales or the same range of physical processes, and their evolution is largely decoupled \citep{poggianti10}. There is also clear evidence that local and global environment do not have the same effect on galaxy evolution. For instance, \cite{vulcani12} found that the local environment has a strong effect on the galaxies' stellar mass function, while the same team  showed that the global environment has no (or much weaker) effect \citep{vulcani13}.

We find that the correlation between $M_*$ and environmental density (Fig.~\ref{fig:envparamt} right panel; Pearson correlation coefficient $0.49$) is much stronger than the $M_*$--\Mvir\ one (Fig~\ref{fig:haloparamt} third panel; correlation coefficient $0.17$), suggesting that the main driver of these correlations is the local density. This is confirmed by Fig.~\ref{fig:haloparamt_bin}. The second panel shows that at fixed density the correlation between $M_*$ and \Mvir\ largely disappears, except, perhaps, for the two highest density bins, although the statistical uncertainties are large. However, the third panel indicates that at fixed \Mvir\ the $M_*$--density relation is still present.  The fourth panel shows that at fixed $M_*$ most of the \Mvir--density correlation vanishes. We conclude that the $M_*$--environment correlations are really driven by the $M_*$--density correlation, while the weaker $M_*$--\Mvir\ correlation is secondary, and it originates on the \Mvir-density and $M_*$--density correlations.  

It could be argued that the detected trend (more massive BGGs live in denser, more massive halos) may be due, at least partially, to a pure statistical effect. If stellar masses are randomly drawn from the mass function of galaxies, massive halos, which host a larger number of galaxies, have a higher probability to host more massive galaxies (see, e.g., \citealt{Tremaine1977,Bhavsar1985,Lin2010,Dobos2011,Paranjape2012,More2012}). However, we argue that this statistical effect cannot be the main driver of the correlation we find. There is quite a lot of evidence indicating that the luminosity of cluster BCGs is inconsistent with just statistical sampling of the cluster galaxy luminosity function: BCGs are generally too bright, and there is too large a gap between the luminosity of the first and second brightest galaxies \citep[][among others]{Sandage1976,Tremaine1977,Bhavsar1985,Dobos2011,More2012,Hearin2013}\footnote{Note, however, that \cite{Paranjape2012} disagree, but \citet{More2012} and \citet{Hearin2013} have argued against their results}. 

If BCGs are not governed by the luminosity/mass function of the rest of the cluster galaxies, the above statistical arguments do not apply. Things may be not so clear for the poorest groups, where the brightest galaxies seem to be compatible with being statistically drawn from the bright end of the galaxy luminosity function, as argued by some of these authors. However, the correlation between BCG mass and environment appears stronger for more massive and denser clusters (see, e.g., rightmost panel of Fig.\ref{fig:envparamt}), where we argue this statistical effect should not apply, and weaker for poorer groups, where the statistical bias should be strongest. If the main driver of the correlation were just the statistical sampling of the luminosity function, we would expect the correlation to be strongest where this effect is most important (low mass and less dense clusters and groups). Since the effect we find is strongest for high-mass and denser clusters, we conclude that the correlation cannot be primarily driven by sampling statistics. 

\vspace{2\normalbaselineskip}

\noindent In summary, in this section we have found that BCGs follow a stellar mass--size relation that is independent of the environment, and that stellar mass is intrinsically correlated with the local environment (or environmental density). In Section~\ref{sec:cDBCGs} we will see how these correlations depend on the morphologies of the BCGs.

\section{Evolutionary History of cD and elliptical BCGs}
\label{sec:cDBCGs}

In \citet{Zhao15} we found that the vast majority of BCGs (over $90$\%) have cD or elliptical morphologies, while only a small minority ($\sim7$\%) are disk galaxies (spirals and S0s), and the remaining few are major mergers. The morphology of these galaxies is clearly linked to their quantitative structural parameters. cDs are generally larger than ellipticals, and their light distributions deviate significantly more from \sersic\ profiles than those of ellipticals. With the additional information presented in this paper we will now explore how morphology and structure are linked to the stellar masses and environments of the BCGs.

\begin{figure*}
\centering{\includegraphics[scale=0.67]{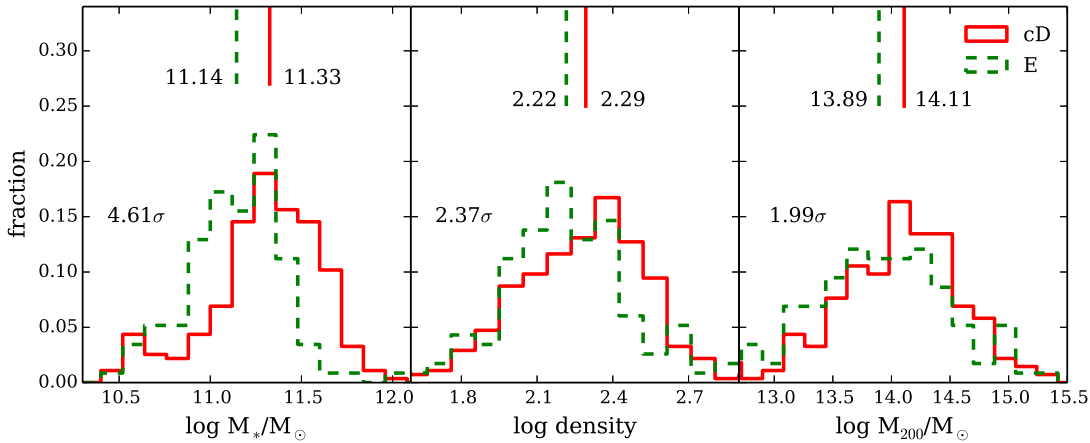}}
\caption{Distribution of $M_*$, environmental density and \Mvir\ for the 275 cD (red solid) and 116 elliptical (green dashed) BCGs in our sample. The $\sigma$ value in each panel indicates the significance (confidence level) of the observed differences between the cD and elliptical BCG parameter distributions. These are derived from two-sample Kolmogorov-Smirnov tests. Statistically, compared with elliptical BCGs, cD galaxies are more massive, tend reside in denser environments, and tend to be hosted by more massive dark matter halos. The median values of the different distributions are indicated by the vertical lines and adjacent numerical values. 
\label{fig:cDEparm}}
%\vspace{-0.2cm}
\end{figure*}

\begin{figure*}
\centering{\includegraphics[scale=0.72]{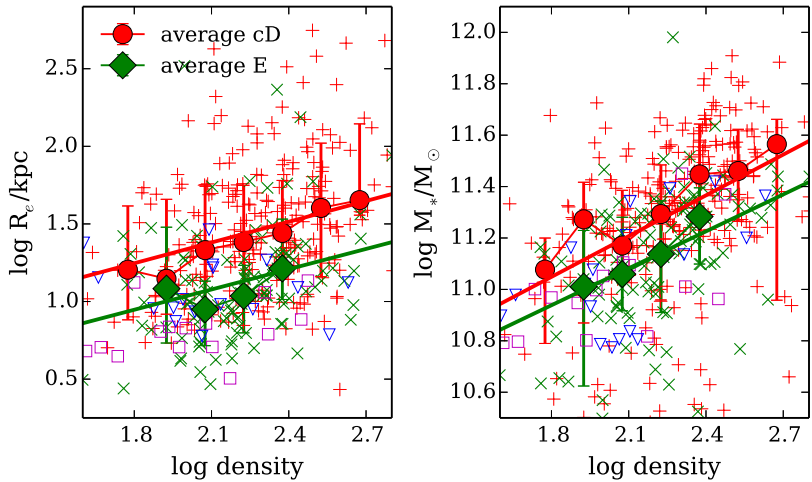}}
\caption{\re--density and $M_{*}$--density relations for BCGs with different morphologies. 
Red plus signs, green crosses, magenta open squares and blue open triangles correspond to cD, elliptical, S0 and spiral BCGs, respectively (as in Fig.~\ref{fig:nrem}). Red filled circles with error bars show the median and the 84 and 16 percentiles for cD galaxies. Green filled diamonds show the same properties for elliptical BCGs. The red and green lines show a linear fit for cD and ellipical BCGs Respectively. It is clear that at the same density, cD galaxies are statistically larger by factor of $\sim 2$ than elliptical BDGs. The stellar mass of cDs is larger by a factor of $\sim 1.4$ than that of ellipticals. Disk BCGs tend to be smaller and less massive. 
\label{fig:cDEbin}}
%\vspace{-0.2cm}
\end{figure*}

In Fig.~\ref{fig:cDEparm} we present the distributions of the stellar masses, environmental densities and parent cluster total masses (\Mvir) for cD and elliptical BCGs.  The left panel clearly shows that cDs have, statistically, larger stellar masses than elliptical BCGs. The median stellar mass of the cDs is $2.1^{+1.7}_{-1.1}\times10^{11}$M$_\odot$, $\sim50$\% larger than that of ellipticals ($1.4^{+0.9}_{-0.6}\times10^{11}$M$_\odot$).  A two-sample Kolmogorov-Smirnov test shows that this difference is significant at the $\sim4.6\sigma$ level. The disk galaxies (not shown in the figure for clarity) are even less massive: the median stellar mass for spirals and S0s is $1.0^{+1.0}_{-0.4}\times10^{11}$M$_\odot$. 

With respect to environmental density (middle panel of Fig.~\ref{fig:cDEparm}), cDs seem to prefer marginally denser regions (by $\sim20$\% on average) than elliptical BCGs, although, statistically, this difference is only significant at the $\sim2.4\sigma$ level. Disk galaxies tend to live in the regions with the smallest densities (a factor of $\sim2$ smaller than cDs). Similarly (right panel of Fig.~\ref{fig:cDEparm}), cDs appear to be hosted by more massive clusters/groups than ellipticals, but once again the difference (a factor of $\sim1.7$ in median \Mvir) is only barely significant ($\sim2\sigma$). 

These differences in the stellar masses and environments of BCGs with different morphologies suggest that their formation histories may be different. In Section~\ref{sec:earlytype} we found that there are intrinsic correlations between \re\ and $M_*$, and between $M_*$ and the environmental density. By exploring the relationship between these properties and the galaxies' morphologies we may be able to shed additional light on the issue of the formation and growth of BCGs. In Fig.~\ref{fig:cDEbin} we show the \re--density relation (left panel) and the $M_*$--density relation (right panel) for cD, elliptical, and disk BCGs. cD and elliptical BCGs show parallel correlations, in the sense that larger and more massive galaxies tend to prefer denser environments.  However, at a fixed environmental density, cDs are, on average, a factor of $\sim2$ larger and $\sim40$\% more massive than elliptical BCGs. Disk galaxies tend to be smaller and less massive, but clear correlations are not seen, perhaps due to the small number statistics.  This correlation is also seen when investigating the relation with the the total mass of the cluster.   These correlations are futhermore certainly due to the fact that there is a different relation between the stellar mass and radius for ellipticals and cD.  This effect is driven by the stellar mass being higher, which then increases the radius.

Note that the observational results presented in this paper, including the differences found between cDs and elliptical BCGs, do not depend on whether the morphological classification is done visually (as shown here) or automatically (based on structural parameter method described in \citealt{Zhao15}). A parallel analysis using the automatic cD/elliptical classification yields entirely consistent results. We are therefore confident that our results are robust, and do not depend significantly on the details of the morphological classification. 

Our empirical results, together with the findings of previous works, suggest a possible scenario linking the evolution of elliptical and cD BCGs. \cite{Whiley08}, \cite{Burke13}, \cite{Burke15} and \cite{Zhang15}, among others, suggest that the stellar mass of BCGs has experienced some (but relatively moderate) growth in the last $\sim6$--$8\,$Gyrs. Although measuring BCG growth is notoriously difficult due to progenitor bias (see \citealt{Shankar2015} for a recent discussion), it seems to be due, mostly, to the effect of minor and major mergers \citep{Burke13}, with minor mergers dominating at later times \citep{Shankar2013,Burke15}. At most, BCGs may have grown by a factor $\sim1.8$ in stellar mass since $z\sim1$, although this factor could have been as small as $\sim1.2$ if about half of the accreted stellar mass from the merging companions became part of the intra-cluster light \citep{Burke15}. This mass growth seems to have been faster in the past, when both minor and major mergers were more common \citep{Burke13}, but these authors also found that BCGs in similar mass clusters can have very different merging histories. Furthermore, \cite{Ascaso11} reported that BCGs have grown in size  by a factor of $\sim 2$ over a similar period. Interestingly, the difference in mass between cDs and elliptical BCGs in similar environments is of the order of $40$\% (i.e., comparable with the measured mass growth), and we find that the difference in size is a factor of $\sim2$ (again, compatible with the measured size growth), but with a very large scatter in both cases. Additionally, \cite{Zhao15} found that, when it could be reliably measured, the fraction of the light (stellar mass) contained in the cD envelopes is of the order of $\sim40$--$60$\%, with significant galaxy-to-galaxy variations.\footnote{Note that the galaxies for which this fraction could be reliable measured are the ones whose profiles are better modelled using two-component \sersic$+$exponential profiles. Since these tend to be the ones with more prominent envelopes, the average fraction of light in cD envelopes is probably closer to $\sim40$\%, the bottom end of the measured range.}  It is therefore plausible that most present-day BCGs started their life as ellipticals, and they subsequently grew, in stellar mass and size, due to mergers to become cDs. In this process, the characteristic cD envelope developed. The large scatter in the stellar masses and sizes of the cDs is explained by their different merger histories. Furthermore, the growth of the BCGs in mass and size seems to be linked to the hierarchical growth of the structures they inhabit: as the groups and clusters become denser and more massive, the BCGs at their centres also grew.

By the present time, most BCGs seem to be well advanced in this process. \cite{Zhao15} found that the majority ($\sim57$\%) of the BCGs are cDs, $\sim21$\% have intermediate cD/E or E/cD morphologies, while ellipticals are a minority ($\sim13\%$). The presence of intermediate morphological classes suggests that this process is still ongoing. Present-day elliptical BCGs may (or may not) develop cD-type envelopes in the future, depending on whether the current merger rate is sufficient. With the limited statistical evidence that we have, we can only speculate  about the origin of the few ($\sim7$\%) BCGs with spiral and S0 morphologies, but perhaps these are the ones which avoided major mergers in their past history and retained their disks. 

If the evolutionary framework we propose is correct, one would expect the morphological mix of BCGs to change with redshift: at earlier times, the fraction of elliptical BCGs should be higher than today, with cDs showing the opposite trend. We have visually examined the images of the 13 BCGs in the ESO Distant Survey \citep{White05} clusters and groups for which deep HST images are available \citep{Desai07}, and morphologically classified them following the same criteria used for the low-redshift sample. The average redshift of these galaxies is $z\sim0.6$. Although cosmological surface-brightness and resolution effects would have to be properly accounted for in a more systematic study, we feel that these HST images have enough resolution and depth (4 orbit exposure) for this purpose. They compare favourably with the SDSS images of the lower-redshift galaxies.  Notwithstanding these possible caveats, we find that 4 of the BCGs are ellipticals, 3 cDs, 4 E/cD or cD/E, one is a spiral, and one is a merger. Although the sample is pitifully small, the trend seems to go in the right direction: the fraction of ellipticals more than doubles when compared with the local sample, while the fraction of cDs halves. There is also a significant fraction of galaxies with intermediate morphologies, suggesting that the transformation process is also happening at these redshifts. Of course, with such small sample, no firm conclusions can be obtained, but at least these findings are compatible with our hypothesis. A systematic study of a large, well-defined sample of BCGs with deep HST images, reaching $z\sim1$, would be required to obtain a definitive answer. 

Numerical simulations and semi-analytic models (see, e.g., \citealt{DeLB07} and references therein) provide a plausible inside-out scenario for the growth of BCGs which is broadly compatible with our findings. At early times ($z\sim1$--$3$), dissipative processes similar to the ones proposed for the formation of normal giant elliptical galaxies were responsible for the building of the BCGs' inner (elliptical-like) stellar component, whose light profile can be well represented by a \sersic\ model. Subsequently, as the structures around BCGs grew hierarchically, the mass and size of these galaxies continued to increase, mainly due to dissipationless (dry) mergers, and the cD envelopes were formed as a result. This picture is also largely consistent with other observations. For example, dry mergers have been directly observed in cluster environments (e.g., \citealt{vanDokkum05}), and it has been suggested that the accreted stars could built up the extended stellar halos observed in BCGs \citep{Abadi06,Murante07}.

\section{Conclusions}
\label{sec:conclude}
Using a large well-defined sample of $425$ nearby Brightest Cluster Galaxies from the catalogue of \cite{Linden07}, we have carried out a study of  the relationships between their internal properties (stellar masses, structural parameters, sizes and morphologies) and their environment. The stellar masses $M_*$ are based on the MPA--JHU SDSS DR7 measurements. The structural parameters (effective radius \re\ and \sersic-index $n$) were derived by \cite{Zhao15} using single \sersic\ profile fits. The visual morphologies were also published by \cite{Zhao15}, who found that the majority ($\sim57$\%) of the BCGs are cDs, $\sim13$\% are ellipticals, $\sim21$\% belong to intermediate cD/E or E/cD classes, and $\sim7$\% have disk morphologies, with spirals and S0s in similar proportions. We use two separate measurements of the environment, the local environmental density \citep{Tempel12}, and the global dark-matter halo virial mass \Mvir\ derived from the cluster velocity dispersions  \citep{Linden07}. Our main conclusions are:

\begin{itemize}
	\item The \sersic-index $n$ does not correlate with the stellar mass $M_*$ or the environment of the galaxies.
	\item The effective radius \re\ of the BCGs correlates with their stellar mass $M_*$, but the scatter is large ($\sim0.3\,$dex in effective radius at a given mass). This correlation does not depend significantly on the environment. 
	\item Almost all BCGs have larger \re\ than non-BCG early-type galaxies of similar $M_*$. The median radius of the BCGs is about twice as large as that of non-BCG early types of similar masses. This difference is largely due to the cD galaxies, which dominate the sample. Moreover, the scatter in the $M_*$--\re\ relation is significantly larger for the BCGs than for the other early-type galaxies, suggesting a more complex formation history.
	\item More massive BCGs tend to inhabit denser regions and more massive clusters, but $M_*$ correlates significantly more strongly with environmental density than with the cluster dark-matter halo mass \Mvir. Indeed, the apparent correlation between $M_*$ and \Mvir\ can be explained by the correlations between \Mvir\ and $M_*$ with environmental density. 
  \item The median stellar mass of cD BCGs is $2.1\times10^{11}$M$_\odot$, $\sim50$\% larger than that of ellipticals ($1.4\times10^{11}$M$_\odot$). BCGs with disk morphologies have even smaller stellar masses (median $1.0\times10^{11}$M$_\odot$).
	\item cDs seem to prefer marginally denser regions (by $\sim20$\% on average) than elliptical BCGs. Disk galaxies tend to live in the regions with the smallest densities. Similarly, cDs appear to be hosted by more massive clusters/groups than ellipticals (by factor of $\sim1.7$ in median \Mvir). However, these differences are only significant at the $2$--$2.4\sigma$ level.
	\item cD and elliptical BCGs show parallel correlations between their stellar masses and environmental densities: larger and more massive galaxies tend to prefer denser environments. However, at a fixed environmental density, cDs are, on average, $\sim40$\% more massive than elliptical BCGs. Due to the correlation between \re\ and $M_*$, cDs and ellipticals also exhibit positive and parallel correlations between their effective radii and the environmental density. cDs are, statistically, twice as large as elliptical BCGs at a given density. Disk BCGs tend to be smaller and less massive.
\end{itemize}

Our results, together with the findings of previous observational and theoretical studies, suggest an evolutionary link between elliptical and cD BCGs. BCGs have experienced a significant growth in mass and size in the last $\sim6$--$8\,$Gyrs, largely due to the effect of minor and major mergers. The mass growth seems to have been faster in the past, when both minor and major mergers were more common, with minor mergers probably playing a dominant role in recent times. The amount of growth in mass and size experienced by BCGs since $z\sim1$ is comparable to the difference in mass and size between cDs and elliptical BCGs in similar environments. Additionally, the fraction of the light (stellar mass) contained in the cD envelopes is also comparable with the average stellar mass difference between cDs and ellipticals. We therefore suggest that most present-day BCGs started their life as ellipticals, and they subsequently grew in stellar mass and size, due to mergers, to become cDs. In this process, the characteristic cD envelope developed. The large scatter in the stellar masses and sizes of the cDs is explained by their different merger histories occurring at $z < 1$. Furthermore, the growth of the BCGs in mass and size seems to be linked to the hierarchical growth of the structures they inhabit: as the groups and clusters became denser and more massive, the BCGs at their centres also grew.

This process is nearing completion by the present time, since the majority of the BCGs in the local Universe have cD morphology. However, the presence of intermediate morphological classes (cD/E and E/cD) suggests that the growth and morphological transformation of some BCGs is still ongoing. It is also possible that today's elliptical BCGs may develop cD-type envelopes in the future, depending on the merger activity they may experience. We also speculate that the BCGs with spiral and S0 morphologies represent the minority of BCGs which avoided major mergers in the past, thus retaining their disks. 

This scenario is broadly compatible with hierarchical inside-out models for the formation and growth of BCGs. Early dissipative processes were responsible for the building of the BCGs' inner elliptical-like stellar component. As the structures around BCGs grew hierarchically, the mass and size of these galaxies continued to increase, mainly due to dissipationless mergers, and the cD envelopes were thus formed.

The evolutionary framework we propose seems to be able to explain the observed properties of BCGs, including the differences between the morphological classes. The obvious next step to test this scenario is to carry out a study of the morphology, mass, structure and environment for a large and statistically robust sample of BCGs as a function of redshift, reaching $z\sim1$. A key piece of evidence would be the evolution of the fraction of cD BCGs with time, and its links with the growth of their masses, sizes and environments.

\section*{Acknowledgments}

DZ's work is supported by a Research Excellence Scholarship from the University of Nottingham and the China Scholarship Council. AAS and CJC acknowledge financial support from the UK Science and Technology Facilities Council. This paper is partially based on SDSS data. Funding for SDSS-III has been provided by the Alfred P. Sloan Foundation, the Participating Institutions, the National Science Foundation, and the U.S. Department of Energy Office of Science. The SDSS-III web site is http://www.sdss3.org/. SDSS-III is managed by the Astrophysical Research Consortium for the Participating Institutions of the SDSS-III Collaboration including the University of Arizona, the Brazilian Participation Group, Brookhaven National Laboratory, Carnegie Mellon University, University of Florida, the French Participation Group, the German Participation Group, Harvard University, the Instituto de Astrofisica de Canarias, the Michigan State/Notre Dame/JINA Participation Group, Johns Hopkins University, Lawrence Berkeley National Laboratory, Max Planck Institute for Astrophysics, Max Planck Institute for Extraterrestrial Physics, New Mexico State University, New York University, Ohio State University, Pennsylvania State University, University of Portsmouth, Princeton University, the Spanish Participation Group, University of Tokyo, University of Utah, Vanderbilt University, University of Virginia, University of Washington, and Yale University. 

%\begin{thebibliography}

\bibliographystyle{mnras}     %plainnat  mn2e
\bibliography{BCG_paper2}

%\end{thebibliography}

\appendix

\section{Comparison with Guo+09}
\label{sec:guo}

\subsection{Structural parameters}
\label{sec:guo1}

\begin{figure}
%\centering{
\hspace{-10pt}\includegraphics[scale=0.54]{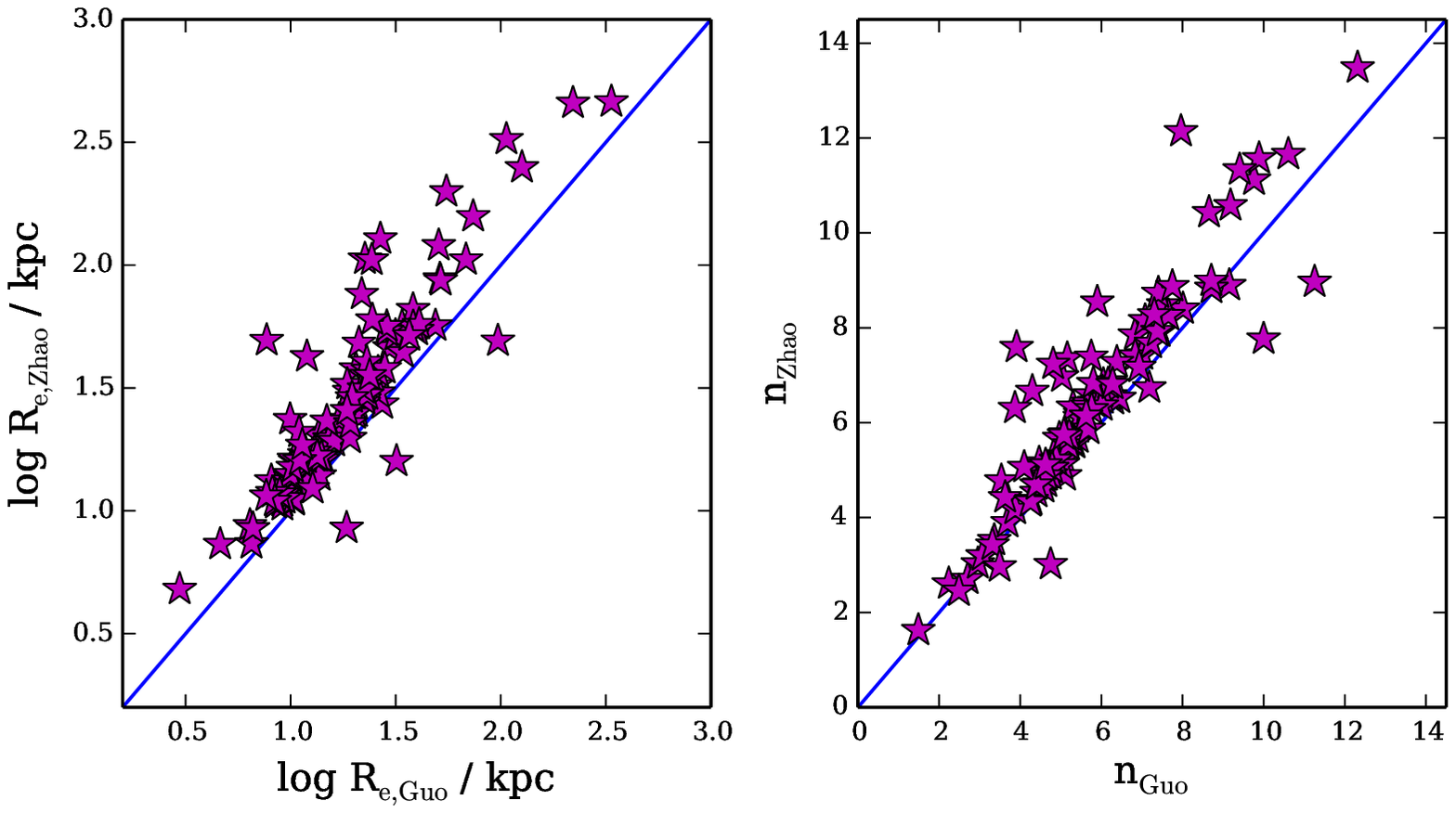}
%}  
\caption{Comparison between the values of the effective radius \re\ and \sersic\ index $n$ obtained by \citet{Zhao15} and \citet{Guo09} for the 104 galaxies in common. The solid lines correspond to the 1-to-1 relation.
\label{fig:ZhaoGuo}}
%\vspace{-0.2cm}
\end{figure}

There are 104 galaxies in common between our sample and that of \cite{Guo09}. A comparison between the measurements of the effective radius \re\ and the \sersic\ index $n$ for these galaxies is presented in Fig.~\ref{fig:ZhaoGuo}. Although the measurements correlate very well, there are some relatively small systematic differences. The median offset between our \re\ measurements and those of \cite{Guo09} is $0.15$dex. The median offset in $n$ is $0.47$. The larger values we obtain are due to improvements in the sky subtraction implemented by \cite{Zhao15}. In that paper we showed that the sky values provided by SDSS DR7 were overestimated due to the presence of extended objects. This is particularly important in crowded fields such as the centres of groups and clusters. We used GALAPAGOS \citep{Barden12} to obtain a more reliable estimate of the sky after removing contamination from neighbouring objects. Although the reduction in the sky values is quite small (typically $\sim0.4$ counts, or 0.3\%), the effect on \re\ and $n$ can be significant for extended objects such as BCGs. More details are provided in section 3.2 of \cite{Zhao15}.      

\subsection{Stellar masses}
\label{sec:guo2}

\begin{figure}
\centering{\includegraphics[scale=0.66]{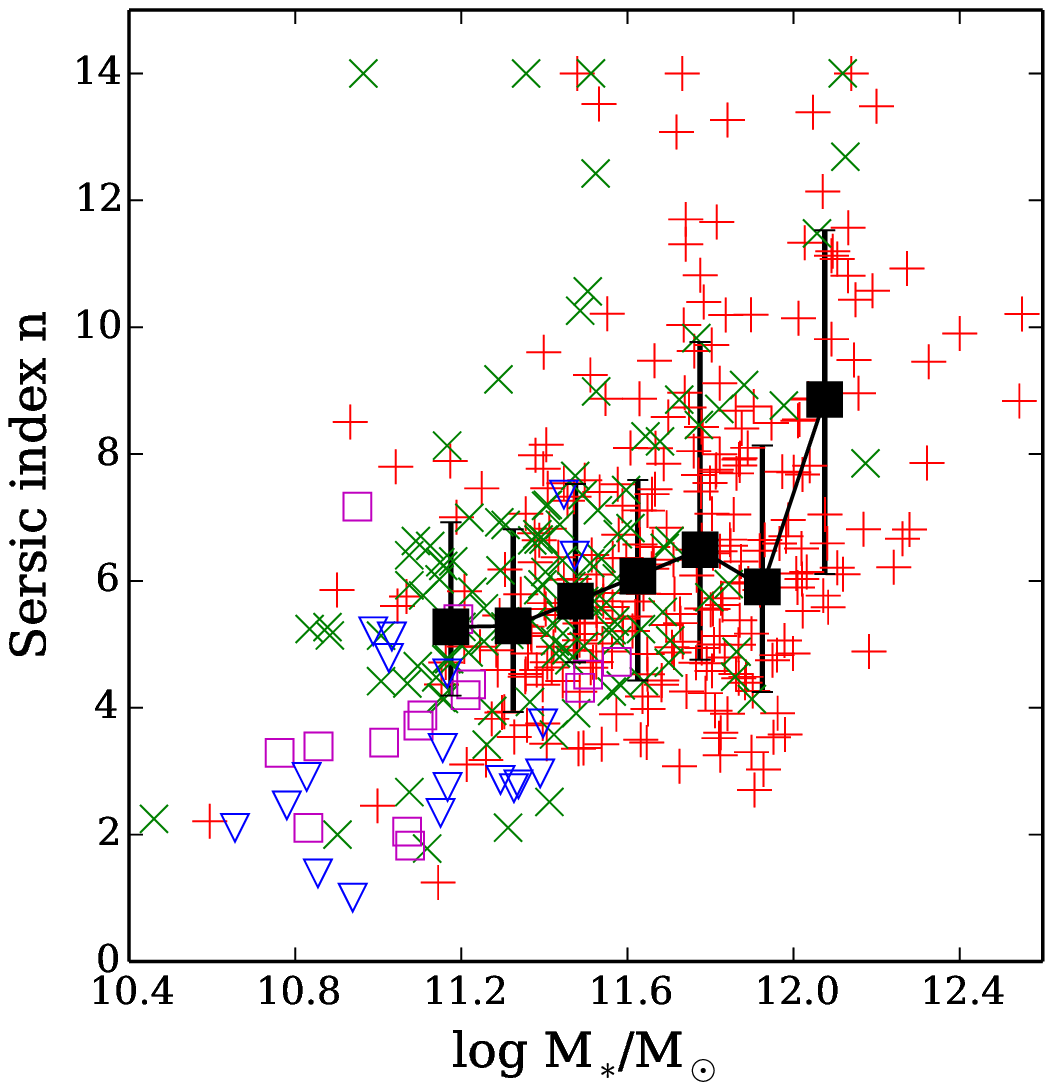}}  
\caption{\sersic\ index $n$ vs. stellar mass for the BCGs in our sample, similar to Fig.~\ref{fig:nrem}, but with the stellar mass $M_*$ is derived following the method described in \citet{Guo09}. Symbols as in Fig.~\ref{fig:nrem}. See text for details.
\label{fig:nmGuo}}
%\vspace{-0.2cm}
\end{figure}

In Section~\ref{sec:stellarmass}, we found no correlation between $n$ and $M_*$ for the BCGs in our sample. This contrasts with the findings of \citet{Guo09}, who show a clear positive correlation in the sense that more massive BCGs seem to have higher values of $n$. In this Appendix we explore the possibility that the correlation found by \citet{Guo09}  may be due to the fact that these authors estimated stellar masses from the total luminosity derived from single \sersic\ model fits. These luminosities (and the derived stellar masses) are therefore model dependent, and, in particular, they will depend on the value of $n$. Since there is a direct relation between the best-fit total flux and $n$ for a \sersic\ profile (see Equations~4 and~6 in \citealt{Peng10}), this dependency could drive the observed correlation. 

In order to confirm this, we have derived stellar masses for the BCGs in our sample following the same method as \citet{Guo09} using our own single \sersic\ fits. Since we have $104$ BCGs in common with \citet{Guo09}, we can check that the values of $M_*$ derived in this way for the galaxies in common agree well with theirs: the scatter in this comparison is below $0.1\,$dex and there is no bias. In Fig.~\ref{fig:nmGuo} we show that, using these model-dependent $M_*$ values, a positive correlation between $n$ and $M_{*}$ is indeed found (Pearson correlation coefficient $0.38$). The correlation we find is qualitatively similar to the one shown in Fig.~6 of \cite{Guo09} when considering the same mass range.  

This indicates that the correlation claimed by \cite{Guo09} may be the consequence of assuming that a \sersic\ model fit provides an accurate representation of the total light distribution of BCGs. This assumption is clearly not correct, particularly for cD galaxies, as demonstrated by previous studies (see \citealt{Zhao15} and references therein). Measuring the total luminosity of a galaxy is far from trivial and, of course, the Petrosian magnitudes used to derive MPA--JHU masses are not without their problems (see, e.g., \citealt{graham05}). We do not claim that the stellar masses we use are better than the ones used by \cite{Guo09}, but they are, at least, model independent and not directly linked to the models used to derive the structural parameters that we study. For these reasons we prefer to use the MPA--JHU masses in this paper. Nevertheless, bearing in mind this uncertainty, we have checked and confirmed that all our conclusions (with the exception of the lack of correlation between $M_*$ and $n$) remain the same if we use \sersic-model based luminosities/stellar masses instead of the MPA--JHU ones.

\label{lastpage}
\end{document}